\documentstyle[epsfig,equation]{aipproc2}
\begin{document}
\def\lsim{\lower2pt\hbox{$\buildrel{<}\over{\sim}$}}
\def\gsim{\lower2pt\hbox{$\buildrel{>}\over{\sim}$}}
\def\rr{{\rm r}}
\def\cc{{\rm c}}
\def\mm{{\rm m}}
\def\ss{{\rm s}}
\newcommand{\FF}{{\cal F}}
\newcommand{\cd}{\cdot}
\newcommand{\ip}{\int_0^{2\pi}}
\newcommand{\al}{\alpha}
\renewcommand{\b}{\beta}
\newcommand{\de}{\delta}
\newcommand{\De}{\Delta}
\newcommand{\ep}{\epsilon}
\newcommand{\ga}{\gamma}
\newcommand{\Ga}{\Gamma}
\newcommand{\ka}{\kappa}
\newcommand{\io}{\iota}
\newcommand{\La}{\Lambda}
\newcommand{\la}{\lambda}
\newcommand{\Om}{\Omega}
\newcommand{\om}{\omega}
\newcommand{\si}{\sigma}
\newcommand{\Si}{\Sigma}
\newcommand{\th}{\theta}
\newcommand{\vth}{\vartheta}
\newcommand{\vph}{\varphi}
\newcommand{\ra}{\rightarrow}
\newcommand{\tr}{\mbox{tr}}
\newcommand{\hor}{\mbox{hor}}
\renewcommand{\baselinestretch}{1.01}
\newcommand{\bea}{\begin{eqnarray}}
\newcommand{\eea}{\end{eqnarray}}
\newcommand{\dd}{\partial}
\def\laq{\raise 0.4ex\hbox{$<$}\kern -0.8em\lower 0.62
ex\hbox{$\sim$}}
\def\double{\baselineskip 24pt \lineskip 10pt}
\newcommand{\sx}{\sigma}
\newcommand{\sei}{\sigma_8}
\newcommand{\sxa}{\sigma_1}
\newcommand{\sxb}{\sigma_2}
\newcommand{\pha}{\phi_1}
\newcommand{\phb}{\phi_2}
\newcommand{\Pha}{\Phi_1}
\newcommand{\Phb}{\Phi_2}
\newcommand{\Phib}{\bar{\Phi}}
\newcommand{\Phab}{\bar{\Phi}_1}
\newcommand{\Phbb}{\bar{\Phi}_2}
\newcommand{\mpl}{m_{Pl}}
\newcommand{\Mpl}{M_{Pl}}
\newcommand{\lx}{\lambda}
\newcommand{\Lx}{\Lambda}
\newcommand{\ex}{\epsilon}
\newcommand{\be}{\begin{equation}}
\newcommand{\ee}{\end{equation}}
\newcommand{\een}{\end{subequations}}
\newcommand{\ben}{\begin{subequations}}
\newcommand{\beq}{\begin{eqalignno}}
\newcommand{\eeq}{\end{eqalignno}}
\def \lta {\mathrel{\vcenter
     {\hbox{$<$}\nointerlineskip\hbox{$\sim$}}}}
\def \gta {\mathrel{\vcenter
     {\hbox{$>$}\nointerlineskip\hbox{$\sim$}}}}
\title{Supersymmetric Hybrid  Inflation:\\
Explaining the Spectrum of Cosmological Perturbations\\
through a Multiple-Stage Inflationary Model} 

\author{Mairi Sakellariadou}
\address{Theory Division, CERN, CH-1211 Geneva 23, Switzerland\\
\vskip 0.2truecm
Centre for Theoretical Physics, University of Sussex\\
Brighton, Falmer, BN1 9QH, United Kingdom}

\maketitle

\begin{abstract}
We explore the possibility that a multiple-stage inflationary scenario 
based on supersymmetric GUT models, can account for the break at 
$k_b \simeq 0.05~h~{\rm Mpc}^{-1}$ in the power spectrum of galaxy 
clustering, while it reproduces the angular power spectrum of cosmic
microwave background anisotropies.
\end{abstract}

\section*{Introduction}

High energy physics may  provide the necessary mechanisms to explain
some of the open questions on the early stages in the history of our
universe.  Conversely, the early universe provides an adequate 
environment for  testing ideas on fundamental physics. A useful
example of the interplay between high energy physics and cosmology is
presented here \cite{mt}.

The inflationary paradigm was proposed in order to explain the
shortcomings of the Big Bang cosmological model. In addition, it
offers a scenario for the generation of the primordial density
perturbations, which can lead to the formation of the observed
large-scale structure. Within the wide variety of inflationary models,
we consider a rather natural candidate, a scenario which arises within
the context of supersymmetric hybrid inflation \cite{cop}.  The
necessary superpotential can arise in supersymmetric GUT models. Since
the observable part of inflation takes place for field values below
the Planck scale, supergravity corrections can be made in terms of
controlled expansions in powers of fields in units of $\mpl$ ($\mpl$
denotes the {\it reduced} Planck scale $\mpl= \Mpl/\sqrt{8 \pi}$,
$\Mpl = 1.22 \times 10^{19}$ GeV).  Such inflationary models
\cite{mecostas} have been shown to survive the supergravity
corrections \cite{megeorge}.

The origin of the measured anisotropies in the cosmic microwave
background (CMB) and the generation and evolution of large-scale
structure in the universe are burning questions of modern
cosmology. The COBE-DMR measurements of the anisotropies of the CMB
lead to the power spectrum of density perturbations at large scales.
This spectrum has to be matched with the power spectrum at small
scales one deduces from galaxy surveys.  The otherwise rather
successful standard cold dark matter (CDM) model with a
scale-invariant initial spectrum of adiabatic perturbations, predicts
too much power at small scales, once normalized to the COBE data at
large scales.  This problem can be resolved within a scenario where
the power spectrum has a break at position indicated by the observed
clustering in the galaxy distribution.  At $k_b \simeq 0.05~h~{\rm
Mpc}^{-1}$ the APM galaxy survey data \cite{data1,peac,subir,brdat2}
indicate a sharp drop of the spectral index $n$, which persists even
after the effects of the non-linear evolution of matter fluctuations
have been removed \cite{brdat2,linear}.

We examine \cite{mt} the possibility of a multiple-stage inflationary
scenario, with a sequence of scales starting near $\mpl$ and
continuing down to the scale implied by COBE. We are interested in
examining whether the COBE data can be made compatible with the data
of galaxy surveys, employing the smallest number of cosmological
parameters.  The COBE-DMR measurements require an inflationary stage
with $n\simeq 1$ and at least 5 e-foldings, while the APM galaxy
survey data support the possibility of a second stage of inflation
with $n \sim 0.6$ that generates $\sim 3$ e-foldings.  The subsequent
$\sim 50$ e-foldings, needed to resolve the horizon and flatness
problems, generate density perturbations at scales that are strongly
affected by non-linear evolution, and therefore can not be constrained
from current observational and experimental data.

\section*{Supersymmetric Hybrid Inflation Model}

Our model \cite{mt} is described by the
superpotential 
\be
W = S_1 \left( -\mu_1^2  
+ \lx_1 \Phab \Pha  + g \Phbb \Phb \right)
+S_2 \left( -\mu_2^2 + \lx_2 \Phbb \Phb \right)~,
\label{two1} 
\ee
where the superfields  $S_1$, $S_2$ are gauge singlets; $S_2$ is the linear 
combination of gauge singlets which does not couple to $\Phab \Pha$ 
\cite{mecostas}. The scalar components of the various superfields
are real. Staying along the $D$-flat directions, we define canonically 
normalized scalar fields according to 
\beq
S_1=\frac{\sxa}{\sqrt{2}}~~~,~~~S_2 =\frac{\sxb}{\sqrt{2}}~~~,~~~
\Pha=\Phab = \frac{\pha}{2}~~~,~~~
\Phb=\Phbb = \frac{\phb}{2}~.
\label{two2} 
\eeq
The potential $V(\sxa,\sxb,\pha,\phb)=\sum_i
\left|\partial W/\partial \Phi_i \right|^2$ is then given by 
\beq
V=&~
\left( \mu_1^2 - \frac{\lx_1}{4} \pha^2 \right)^2
-\frac{g}{2}\mu_1^2\phb^2 +\frac{g^2}{16}\phb^4 
+\frac{g\lx_1}{8} \pha^2 \phb^2 
+ \frac{\lx_1^2}{4}\sxa^2 \pha^2 
\nonumber \\
&
+ \left( \mu_2^2 - \frac{\lx_2}{4} \phb^2 \right)^2
+ \frac{1}{4}\left(g\sxa+\lx_2\sxb\right)^2 \phb^2~,
\label{two3} 
\eeq
where the mass scales $\mu_1$, $\mu_2$ are chosen 
to satisfy the inequality $\mu_1 > \mu_2$.

During the first stage of inflation $\pha=\phb=0$, and the vacuum
energy density is $V=\mu_1^4+\mu_2^4$ independent of $\sx_{1,2}$.
Along this flat direction, 
the mass terms of the $\phi_{1,2}$ fields are
\beq
M^2_{\pha}=
-\lx_{1} \mu_{1}^2 + \frac{\lx_{1}^2}{2} \sxa^2~~~,~~~
M^2_{\phb}=
-\lx_{2} \mu_{2}^2 - g \mu_1^2
+ \frac{1}{2} \left(g \sxa +\lx_2 \sxb \right)^2~.
\label{two5b} 
\eeq
The mass term $M^2_{\pha}$ becomes negative for 
$\sxa^2  < \sx^2_{1ins} = 2 \mu_1^2/\lx_1$, indicating the presence of an 
instability which can lead to the growth of the $\phi_{1}$ field.

During the second stage of inflation, $\sxa=0$, $\pha^2= 4\mu_1^2/\lx_1$, 
$\phb=0$; the vacuum energy density is $V=\mu_2^4$, independent of $\sxb$. 
Along this flat direction, the mass term of the $\phb$ field is 
\be
M^2_{\phb}=~
-\lx_{2} \mu_{2}^2 + \frac{\lx_{2}^2}{2} \sxb^2~.
\label{two6a} 
\ee
An instability appears for $\sxb^2  < \sx^2_{2ins} = 2 \mu_2^2/\lx_2$,
which can lead to the growth of the $\phi_2$ field.

The flatness of the potential is lifted by radiative corrections.
During the first stage of inflation and for $\sx_{1,2}$ far above the 
instability points, the one-loop contribution to the effective potential, 
assuming that the coupling $g$ is sufficiently small, is
\beq
\Delta V(\sxa,\sxb) \simeq~ 
\frac{M_1}{16 \pi^2}  \lx_1^2 \mu_1^4
\left[ \ln\left(
\frac{\lx_1^2 \sxa^2}{2 \Lambda_1^2} 
\right) 
+ \frac{3}{2}
\right]
+ \frac{M_2 }{16 \pi^2}  
\left( \lx_2 \mu_2^2 + g \mu_1^2\right)^2
\left[ \ln\left(
\frac{\left(g\sxa + \lx_2 \sxb\right)^2}{2 \Lambda_2^2} 
\right) 
+ \frac{3}{2}
\right]~,
\label{two7} 
\eeq 
where $M_{1,2}$ are the dimensionalities of the representations
of the groups to which the superfields $\Phi_{1,2}$ belong.  The above
contribution provides the slope that leads to the slow rolling of the
$\sx_{1,2}$ fields during the first stage of inflation.

During the second stage of inflation the slope for the $\sxb$ field is 
provided by the radiative correction
\be
\Delta V(\sxa,\sxb) \simeq 
\frac{M_2}{16 \pi^2}   \lx_2^2 \mu_2^4
\left[ \ln\left(
\frac{\lx_2^2 \sxb^2}{2 \Lambda_1^2} 
\right) 
+ \frac{3}{2}
\right]~.
\label{two8} 
\ee
Based on the existence of these two flat directions, with a small slope 
generated by logarithmic radiative corrections, we will construct \cite{mt}
a model for a multiple-stage inflation, in the context of supersymmetry
and supergravity.

\section*{The Inflationary Stages}

The first stage of inflation starts at $t_{1i}$ and lasts until
$t_{1f}$, when $\sxa^2\left(t_{1f}\right) \equiv \sx_{1f}^2\simeq 
M_1 \lx^2_1 \mpl^2 /(8 \pi^2)\ll \sx^2_{1ins}$ and the ``slow-roll''
conditions for $\sxa$ cease to be satisfied. The Hubble parameter during this 
period $H^2_1 \simeq (\mu_1^4+\mu^4_2)/(3 \mpl^2)$ is almost constant, and the
total number of e-foldings is \cite{mt} 
\be
N_{1tot}
=\frac{4\pi^2}{M_1\lx_1^2} \frac{\mu_1^4+\mu^4_2}{\mu^4_1}  
\frac{\sx_{1i}^2-\sx_{1f}^2}{\mpl^2}~,
\label{three4} \ee
where $\sx_{1i}^2 \equiv \sxa^2\left(t_{1i}\right)$.  We can choose
the parameters in our model \cite{mt}, so that the evolution of $\sxb$
during the first stage of inflation can be neglected.

Then it follows an intermediate stage, lasting  between times $t_{1f}$
and $t_{2i}$, during which the scale factor increases by \cite{mt}
\be
N_{int}=\frac{2}{3(1+w)} \ln \left(
\frac{H_1}{H_2} \right)~,
\label{la} 
\ee
where $w$ characterizes the equation of the state of the universe 
($p=w \rho$). 
Assuming the massive fields $\sxa$, $\pha$ have fast decay channels into 
lighter species that eventually thermalize, the fields $\sxb$, $\phb$ 
remain constant during the intermediate stage and the universe is in the 
radiation-dominated era \cite{mt}.

At time $t_{2i}$ the second inflationary stage starts and lasts until
$t_{2f}$, when $\sxb^2\left(t_{2f}\right) \equiv \sx_{2f}^2 \simeq 
M_2 \lx^2_2 \mpl^2 /(8 \pi^2) \ll \sx^2_{2ins}$
and the ``slow-roll''conditions for $\sxb$ cease to be satisfied. The Hubble 
parameter during this period $H^2_2 \simeq \mu_2^4/(3 \mpl^2)$ is almost 
constant, and the total number of e-foldings is \cite{mt}
\be
N_2(t)=\frac{4\pi^2}{M_2\lx_2^2} 
\frac{\sx_{2i}^2-\sx_{2f}^2}{\mpl^2},
\label{three9} \ee
where $\sx_{2i}^2 \equiv \sxb^2\left(t_{2i}\right)$.

Subsequent inflationary stages are needed in order to resolve the
flatness and horizon problems of standard cosmology. We can easily
extend our model \cite{mt} to accommodate additional inflationary
stages which can provide the remaining required number of $\sim 50$
e-foldings.  Choosing the parameters of our model, the above
discussion of the first two inflationary stages remains unchanged.

\section*{The Primordial Spectrum}

The primordial spectrum of density inhomogeneities has its origin in
the quantum fluctuations that crossed outside the horizon during
inflation \cite{freeze}.  For $k\lta k_2=a_{2i}H_2$, or $k \gta
k_1=a_{1f}H_1$, the Hubble radius is crossed only once, during the
first, or second, stage of inflation respectively.  On the other hand,
the scales $k_2<k < k_1$ cross the Hubble radius three times: they
exit the horizon during the first stage, re-enter during the
intermediate stage and exit again during the second stage of inflation.

Assuming that the field fluctuations are random gaussian variables,
the spectrum of adiabatic density perturbations during the first stage
of inflation is \cite{mt}
\be
\delta_H(k) \simeq
\frac{8\pi}{5M_2}
\left( \frac{\mu_2^2}{\lx_2 \mu_2^2+g \mu_1^2} \right)^2
\frac{H_1\sx_{2i}}{\mpl^2}~.
\label{four10} \ee
The predicted spectrum is scale invariant with a spectral index
$n=1$ to a very good accuracy \cite{mt}.

The spectrum of adiabatic density perturbations during the second stage
of inflation is \cite{mt}
\be
\delta_H(k) \simeq
\frac{8\pi}{5M_2 \lx_2^2}
\frac{H_2\left[
\sxb
 \right]_{k=aH_2}}{\mpl^2}~,
\label{four11} 
\ee
while the spectral index is given by \cite{mt}
\be
n-1 =-\frac{M_2 \lx_2^2}{4 \pi^2}  \frac{\mpl^2}{\sxb^2(k)}\neq 1~.
\ee
We believe \cite{mt} that for scales $k_2<k<k_1$ there is a smooth 
interpolation between the two parts of the spectrum.

\section*{Comparison with Observations}

We compare the predictions of our model with the experimental data
from COBE and the observational data from the APM galaxy survey.  We
consider \cite{mt} a CDM model with $\Omega_{\Lx}=0$, $h=0.5~~(H_0
\equiv h~100~{\rm km/s/Mpc})$, and $\Omega_{matter}=1$.  We assume
$\Omega_{b}=0.05$, $\Omega_{CDM}=0.95$, $Y_{He}=0.24$, while we do not
consider any massive neutrinos. The values of the parameters for our
model are \cite{mt}: $\lx_1=1$, $\lx_2=0.1$, $g=7.1 \times 10^{-3}$,
$\mu_1/\mpl=3.9 \times 10^{-3}$, $\mu_2/\mpl=8.8 \times 10^{-4}$,
$\sx_{1i}/\mpl=0.36$, $\sx_{2i}/\mpl=0.025$.  We obtain $5$ e-foldings
with an almost scale-invariant spectrum of perturbations for the first
stage of inflation, and $3$ e-foldings with a spectrum having a
spectral index $n \simeq 0.6$ for the second stage.  The intermediate
stage affects the scales $k_2<k<k_1$ with $k_1/k_2 \simeq 4.4$.  The
values of the power spectrum for $k_2$ and $k_1$ are determined by
Eqs.~(\ref{four10}), (\ref{four11}) and satisfy $P(k_2)/P(k_1)\simeq
11.8$.

The primordial spectrum of fluctuations arising in our model 
generates matter and radiation perturbations, which, after amplification,  
give rise to the observed large scale structure and the measured 
anisotropies of the CMB. Thus, to test our model, we confront its
theoretical predictions with the measured anisotropies
of CMB and the observed clustering in the galaxy distribution. 
For the comparison with the CMB data, we assume \cite{mt} a form of
the power spectrum in the range $k_2<k<k_1$ that interpolates
smoothly between the spectrum in the ranges $k>k_1$ and $k<k_2$.

In order to calculate the predictions of our model for the angular power 
spectrum of CMB anisotropies, we have used \cite{mt} the code 
CMBFAST of Seljak and Zaldarriaga \cite{cmbfast}.  Fig.~1 illustrates 
\cite{mt} our theoretical predictions for the angular power spectrum of 
CMB anisotropies against the most recent CMB flat-band power measurements.

\begin{figure}[htb]
\centerline{\epsfig{file=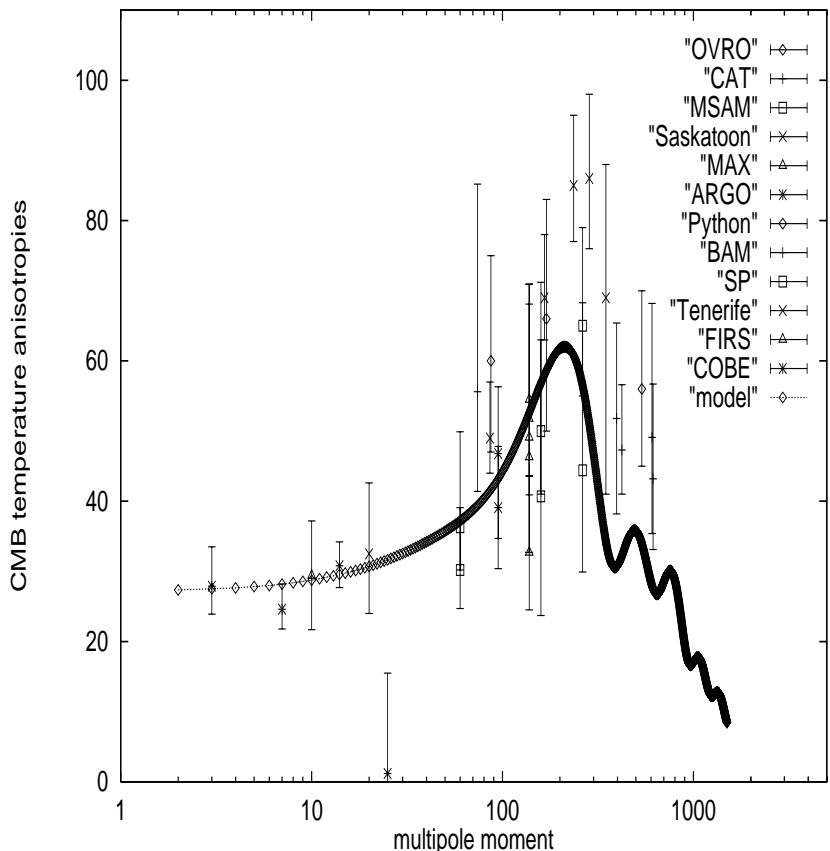,height=4.5in,width=4.5in}}
\caption{Theoretical predictions of our model against the
most recent CMB flat-band power measurements. }
\label{fig1}
\end{figure}

The predicted first acoustic peak in our multiple-stage model
\cite{mt}, is similar to that of the standard CDM (sCDM) model, with a
scale invariant spectrum at all scales.  Only for multipoles $l \sim
500-1000$, the predictions of our model differ significantly of those
deduced from the sCDM model. More precisely, a drop of the values of
$C_l$ is observed relative to the sCDM model. This was expected since
during the first inflationary stage the spectrum is scale invariant,
while the intermediate and second stages affect only the highest
multipoles. Better agreement with the data is obtained \cite{mt} for a
CDM model with $\Omega_{\Lx}=0.5$, $h=0.5$, $\Omega_{b}=0.05$ and
$\Omega_{CDM}=0.45$ within our multiple-stage inflationary scenario.

We then turn to the comparison of our theoretical predictions
\cite{mt} with the power spectrum of galaxy clustering derived from
the angular APM galaxy survey \cite{apm}.  In Fig.~2 we
plot \cite{mt} the power spectrum generated in our multiple-stage model
\cite{mt}, together with the APM power spectrum and the linear
spectrum as predicted by Refs.~\cite{brdat2,linear}.
The spectrum is flat for $k \leq 0.06 ~h{\rm Mpc}^{-1}$ and
tilted with $n=0.6$ for $k \geq 0.26 ~h{\rm Mpc}^{-1}$.

\begin{figure}[htb]
\centerline{\epsfig{file=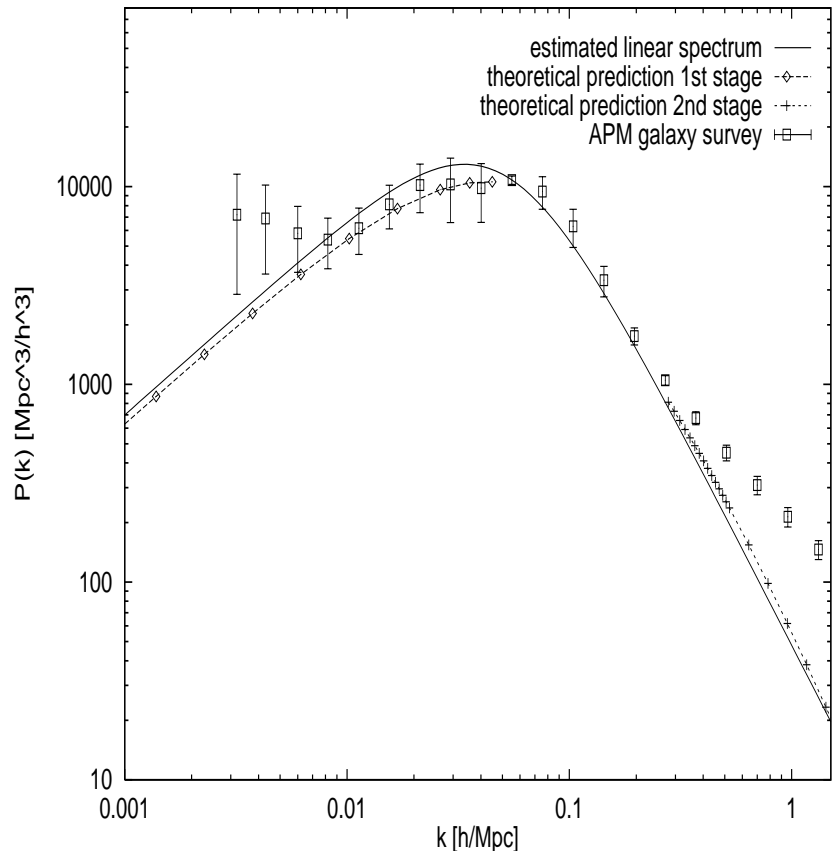,height=4.5in,width=4.5in}}
\caption{Theoretical power spectrum  together with the APM power spectrum,
and linear spectrum.}
\label{fig2}
\end{figure}

Finally, calculating the parameter $\sei$, we find \cite{mt} $\sei$
near 0.75, both for $\Omega_\Lx=0$ and $\Omega_\Lx=0.5$, which is in
good agreement with the values deduced from the abundances of rich
clusters of galaxies: $\sigma_8\simeq 0.6 \pm 0.2$ for
$\Omega_{matter}=1$, $\Omega_\Lx=0$ \cite{wef93}, and $\sigma_8\simeq
0.85 \pm 0.3$ for $\Omega_{matter}=0.5$, $\Omega_\Lx=0.5$
\cite{viana}.

\section*{Conclusions}

In the context of supersymmetric hybrid inflation, we presented
\cite{mt} a multiple-stage inflationary scenario which could account
for some of the characteristics of the  current observational and
experimental data. We proposed a rather natural mechanism
to  generate a primordial spectrum of adiabatic
density perturbations with a break at $k_b \simeq 0.05~h~{\rm
Mpc}^{-1}$. The existence of this break in the power spectrum of
galaxy clustering was suggested by the angular APM galaxy survey. In
addition, our model reproduces successfully the angular power spectrum
of CMB anisotropies.

More precisely, we described a scenario of multiple short bursts of
inflation.  The first two observable inflationary stages, for which we
have constraints from current observational and experimental data,
generate $\sim 8$ e-foldings. The subsequent
stages, providing the required additional $\sim 50$ e-foldings,
generate perturbations at scales that are strongly affected by
non-linear evolution. It is thus very hard to extract any relevant
information. The first inflationary stage generates $\simeq 5$
e-foldings and a scale invariant spectrum for scales $k \lta
0.06~h~{\rm Mpc}^{-1}$. The second inflationary stage generates
$\simeq 3$ e-foldings and a spectrum with spectral index $n \simeq
0.6$ for scales $k \gta 0.26~h~{\rm Mpc}^{-1}$. Between these two
inflationary stages there is a rather complicated intermediate stage
of normal expansion, and therefore we are unable to calculate the
spectrum for $0.06~h~{\rm Mpc}^{-1} \lta k \lta 0.26~h~{\rm
Mpc}^{-1}$. However, we believe that in this intermediate stage there
is no significant new feature and we can interpolate smoothly between
the two parts of the spectrum that we have computed.

\section*{Acknowledgements}
It is a pleasure to thank the organizers of the {\it 3k Cosmology} meeting, 
and in particular, Francesco Melchiorri, for inviting me to present 
this paper. 
The work discussed here is based on a paper I wrote with Nikos Tetradis.

\end{document}